\documentclass{elsart1p}
\usepackage{graphicx}
\usepackage{amssymb}

\usepackage{psfrag}

\psfrag{Q2}{{\tiny $Q^2$}}

\psfrag{EpGd}{{\scriptsize $E^{\pi^0 p}_{0+}/G_D$}}

\psfrag{LpGd}{{\scriptsize $L^{\pi^0 p}_{0+}/G_D$}}

\psfrag{EnGd}{{\scriptsize $E^{\pi^+ n}_{0+}/G_D$}}

\psfrag{LnGd}{{\scriptsize $L^{\pi^+ n}_{0+}/G_D$}}

\newcommand{\beq}[1]{
\begin{equation}\label{#1}}
\newcommand{\eeq}{\end{equation}}
\newcommand{\bea}[1]{
\begin{eqnarray}\label{#1}}
\newcommand{\eea}{\end{eqnarray}}

\begin{document}
\begin{frontmatter}
%
%
%
\title{Threshold Pion Production at Large Momentum Transfers}
%
%
\author[label1]{V.~M.~Braun},
\author[label1,label2]{D.~Yu.~Ivanov},
\author[label1]{A.~Lenz},
\author[label1]{A.~Peters}
  \address[label1]{Institut f\"ur Theoretische Physik, Universit\"at
          Regensburg, D-93040 Regensburg, Germany}
  \address[label2]{Sobolev Institute of Mathematics, 630090 Novosibirsk, Russia}

%
\begin{abstract}
We consider threshold pion electroproduction on a proton target
for photon virtualities in the region $1-10$~GeV$^2$.
The S-wave multipoles at threshold, $E_{0+}$ and $L_{0+}$, are calculated 
using light-cone sum rules. 
\end{abstract}
\end{frontmatter}
%
%
Pion electroproduction at threshold from a proton target,
$e(l)+p(P) \to e(l') + \pi^+(k) + n(P')$ 
and
$e(l)+p(P) \to e(l') + \pi^0(k) + p(P')$,
can be described in terms of two generalised form factors \cite{Braun:2006td,Braun:2007pz}
\begin{eqnarray}
\label{def:G12}
\lefteqn{
 \langle N(P')\pi(k) |j_\mu^{em}(0)| p(P)\rangle=}
\\ &=& 
- \frac{i}{f_\pi} \bar N(P')\gamma_5
  \left\{\left(\gamma_\mu q^2 - q_\mu \!\not\! q\right) \frac{1}{m^2} G_1^{\pi N}(Q^2)
    - \frac{i \sigma_{\mu\nu}q^\nu}{2m} G_2^{\pi N}(Q^2)\right\}N(P)\,,
\nonumber
\end{eqnarray}
which can be related to the S-wave transverse $E_{0+}$ and longitudinal $L_{0+}$ multipoles
(at threshold). Here and below $m=939$~MeV is the nucleon mass.

The celebrated low-energy theorem (LET) \cite{KR,Nambu:1997wa,Nambu:1997wb} relates the S-wave multipoles or, equivalently,
the form factors $G_1,G_2$ at threshold, to the nucleon electromagnetic and axial form factors in the chiral limit 
$m_\pi=0$
\begin{eqnarray}
 \frac{Q^2}{m^2} G_1^{\pi^0 p} &=& \frac{g_A}{2}\frac{Q^2}{(Q^2+2m^2)} G_M^p\,,
\label{LET}
\quad  G_2^{\pi^0 p} =  \frac{2 g_A m^2}{(Q^2+2m^2)} G_E^p\,,
\\
  \frac{Q^2}{m^2} G_1^{\pi^+ n} &=& \frac{g_A}{\sqrt{2}} \frac{Q^2}{(Q^2+2m^2)} G_M^n + \frac{1}{\sqrt{2}}G_A\,,
 \quad G_2^{\pi^+ n} =   \frac{2\sqrt{2} g_A m^2}{(Q^2+2m^2)} G_E^n\,.
\nonumber
\end{eqnarray} 
Here the terms in $G_{M,E}$ are due to pion emission off the initial proton state, 
whereas for charged pion in addition there is a contribution corresponding
to the chiral rotation of the electromagnetic current \cite{KR}. 

The subsequent discussion concentrated mainly on the corrections to (\ref{LET}) due to finite pion mass 
\cite{Vainshtein:1972ih,Scherer:1991cy}.
More recently, the threshold pion production for small $Q^2$ was reconsidered and the low-energy theorems re-derived 
in the framework of the chiral perturbation theory (CHPT), see \cite{Bernard:1995dp} for a review. The new insight gained from 
CHPT calculations \cite{Bernard:1992ys} is that the expansion at small $Q^2$ has to be done with care as the limits 
$m_\pi\to0$ and $Q^2\to 0$ do not commute, in general. The LET predictions seem to be in good agreement with experimental 
data on pion photoproduction \cite{Drechsel:1992pn}, However, it appears \cite{Bernard:1992rf,Bernard:1995dp} 
that the S-wave electroproduction cross section for already 
$Q^2 \sim 0.1$~GeV$^2$ cannot be explained without taking into account chiral loops.   
    
{}For larger momentum transfers the situation is much less studied as the power counting of CHPT cannot be applied.
The traditional derivation of LET using PCAC and current algebra does not seem to be affected as long as
the emitted pion is  'soft' with respect to the initial and final state nucleons simultaneously.
The corresponding condition is, parametrically, $Q^2 \ll \Lambda^3/m_\pi$ (see, e.g. \cite{Vainshtein:1972ih})
where $\Lambda$ is some hadronic scale, and might be satisfied for $Q^2\sim 1$~GeV$^2$ or even higher.  
We are not aware of any dedicated analysis of the threshold production in the $Q^2\sim 1$~GeV$^2$ region, 
however.

{}It was suggested \cite{PPS01} that in the limit of very large momentum transfers 
the standard pQCD collinear factorisation approach becomes applicable and the 
helicity-conserving $G_1^{\pi N}$ form factor can be calculated for $m_\pi=0$ 
in terms of chirally rotated nucleon distribution amplitudes. 
In practice one expects that the onset of the pQCD regime is postponed
to very large momentum transfers because the factorisable contribution involves a small factor $\alpha_s^2(Q)/\pi^2$ 
and has to win over nonperturbative ``soft'' contributions that are suppressed by an extra power of $Q^2$ but do not 
involve small coefficients.    

The purpose of this study is to suggest a realistic QCD-motivated model for the $Q^2$ dependence of the $G_{1,2}$ 
form factors alias S-wave multipoles at threshold in the region $Q^2 \sim 1-10$~GeV$^2$ that can be accessible
in current and future experiments in Jefferson Laboratory and elsewhere (HERMES, MAMI).    
In Ref.~\cite{Braun:2001tj} we have developed a technique  to calculate baryon form factors
for moderately large $Q^2$ using light-cone  sum rules (LCSR).  
This approach is attractive because in LCSRs  ``soft'' contributions to the form factors are calculated in 
terms of the same nucleon distribution amplitudes (DAs) that
enter the pQCD calculation and there is no double counting. Thus, the LCSRs provide one with the most 
direct relation of the hadron form factors and distribution amplitudes that is available at present, 
with no other nonperturbative parameters. 
The same technique can be applied to pion electroproduction, taking into account the
semi-disconnected pion-nucleon contributions in the intermediate state. 
In Refs.~\cite{Braun:2006td,Braun:2007pz} the $G_1$ and  $G_2$ form factors are estimated in the 
LCSR approach for the range of momentum transfers $Q^2 \sim 1-10$~GeV$^2$. 
We demonstrate that the LET results in (\ref{LET}) are indeed
reproduced at $Q^2\sim 1$~GeV$^2$ to the required accuracy $\mathcal{O}(m_\pi)$, whereas the pQCD 
contribution considered in \cite{PPS01} formally corresponds to the leading (at large $Q^2$) part
of the NNLO radiative correction $\sim {\mathcal O}(\alpha_s^2)$ to the sum rules.
Hence our approach describes both high-$Q^2$ and low-$Q^2$ limits correctly and presents an 
extrapolation in between that makes maximal use of quark-hadron duality and dispersion relations.

\begin{figure}[t]
\begin{center}
  \includegraphics[width=0.85\textwidth,angle=0]{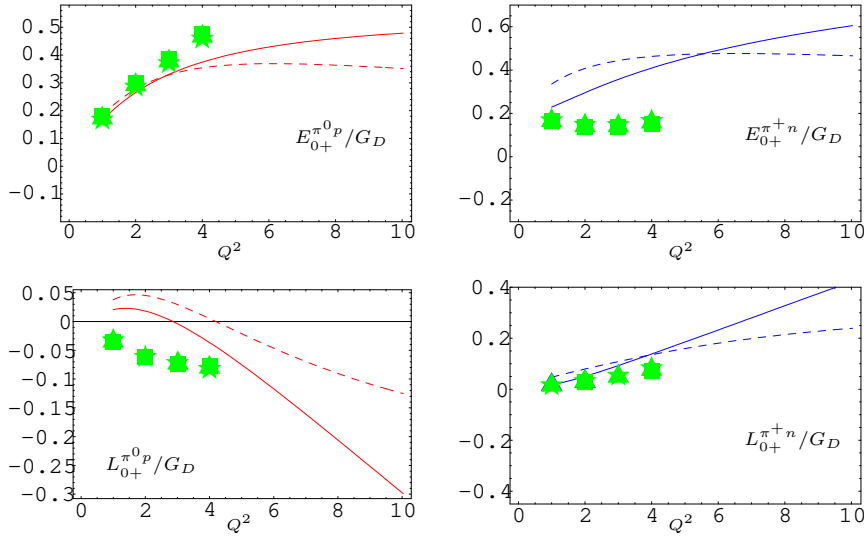}
\end{center}
\caption{The tree-level LCSR results (curves) compared to MAID07 
\cite{Drechsel:2007if} (points) 
for the $Q^2$ dependence of the electric and longitudinal 
partial waves at threshold, $E_{0+}$ and $L_{0+}$ (in units of GeV$^{-1}$),  
normalised to the dipole formula.  
}
\label{fig:G12}
\end{figure}

Accurate quantitative predictions are difficult for several reasons, e,g, because the nucleon distribution amplitudes
are poorly known. In order to minimise the dependence of various parameters one may use 
the LCSRs to predict certain form factor ratios only, and then normalise to the electromagnetic nucleon form factors 
as measured in the experiment, see \cite{Braun:2007pz} for the details.
The results are shown by the solid curves in Fig.~\ref{fig:G12}, where the four partial waves at 
threshold 
are plotted as a function of $Q^2$, normalised to the dipole formula
$
   G_D(Q^2) = 1/(1+Q^2/\mu_0^2)^2
$
where $\mu_0^2 = 0.71$ GeV$^2$. 
To give a rough idea about possible uncertainties, 
the ``pure'' LCSR predictions (all form factors and other input taken from the sum rules) 
are shown  by dashed curves for comparison.  
The accuracy can be improved in future and requires calculation of radiative corrections
to the LCSRs, especially if sufficiently precise
lattice calculations of the moments of nucleon 
distribution amplitudes become available.

%
%
%

%
\end{document}